\documentclass[prl,amsmath,amssymb,superscriptaddress,letterpaper]{revtex4}

\usepackage{graphicx}
\usepackage{bm}
\usepackage{subfigure}

\begin{document}

\title{Maximally Informative Stimuli and Tuning Curves for Sigmoidal Rate-Coding Neurons and Populations}

\author{Mark D. McDonnell}
 \email{mark.mcdonnell@unisa.edu.au}
\affiliation{Institute for Telecommunications Research, University of South Australia, SA 5095, Australia}
\author{Nigel G. Stocks}
 \email{n.g.stocks@warwick.ac.uk}
\affiliation{School of Engineering, University of Warwick, Coventry CV4 7AL, United Kingdom}
\date{\today}

\begin{abstract}
A general method for deriving maximally informative sigmoidal tuning curves for neural systems with small normalized variability is presented. The optimal tuning curve is a nonlinear function of the cumulative distribution function of the stimulus and depends on the mean-variance relationship of the neural system. The derivation is based on a known relationship between Shannon's mutual information and Fisher information, and the optimality of Jeffrey's prior. It relies on the existence of closed-form solutions to the converse problem of optimizing the stimulus distribution for a given tuning curve.  It is shown that maximum mutual information corresponds to constant Fisher information only if the stimulus is uniformly distributed. As an example, the case of sub-Poisson binomial firing statistics is analyzed in detail.
\end{abstract}

\pacs{87.19.lc,87.19.lo,87.19.ls,87.19.lt}

\maketitle

Stimuli transduced by biological sensory systems are communicated to
the brain by short duration electrical pulses known as {\em action
potentials}~\cite{Dayan,Gerstner}. These `spikes' are generated
by synaptic transmission from receptor cells, and propagate to the brain along nerve
fibers.

The derivation in this paper applies to {\em rate coding} neurons or neural populations. Although the results may be relevant for cortical neurons, they are more likely to be useful for sensory neuronal populations whose function is to code a random and continuously varying stimulus parameter, and where the variability between neurons is largely uncorrelated, e.g. fibres of the cochlear nerve~\cite{Johnson.76}.

In rate coding neurons individual action potential timings are not
important, and information is coded by {\em mean firing rate}~\cite{Dayan,Gerstner}, i.e. the average number of action potentials
observed while a stimulus $x$ is constant for
some duration $t$. Experimentally, if firing rate measurements are obtained for a range of stimulus
intensities, an average {\em tuning curve} (also variously known as the stimulus-response
curve, gain function or rate-level function) can be plotted as a function of the stimulus
intensity~\cite{Dayan,Butts.06,Salinas.06}.

There is usually natural variability in the firing rate for a fixed stimulus, which often is called {\em noise}~\cite{Dayan,Gerstner}. Although this variability has led to many previous
Shannon information theoretic~\cite{Cover2} studies of neurons and population of neurons,
e.g.~\cite{Stein.67,Stemmler.96,Kang.01,Bethge.03,Hoch.03a}, results for the tuning curve that maximizes information transfer for a rate-coding neuron appear less frequently~\cite{Nadal.94,Brunel.98,Butts.06,Salinas.06}. Furthermore, such studies usually focus on neurons that have a so-called {\em preferred stimulus}, and a unimodal tuning curve.

\begin{figure}[t]
\begin{center}
{\subfigure[~Stimulus PDFs compared with $V^o(x)$ and ${T^o}'(x)$.]{\includegraphics[scale=0.8]{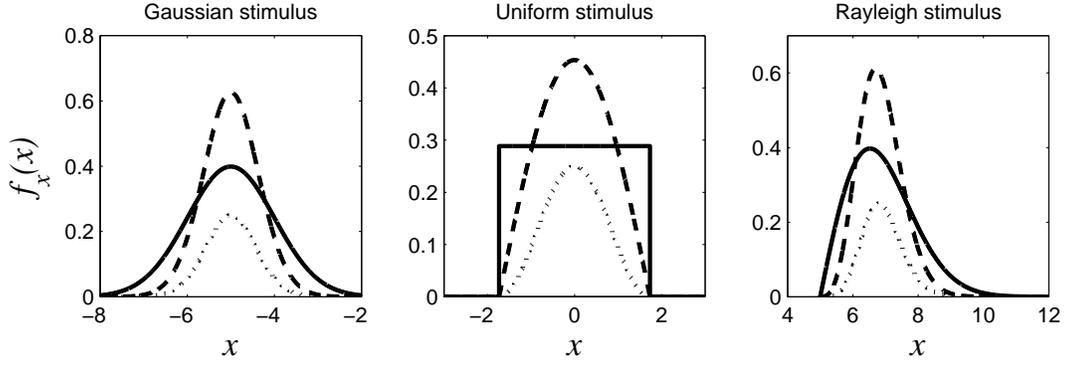}\label{f:varTuning}}
\subfigure[~Three optimal tuning curve $T^o(x)$ for $f_x(x)$ in (a).
]{\includegraphics[scale=0.8]{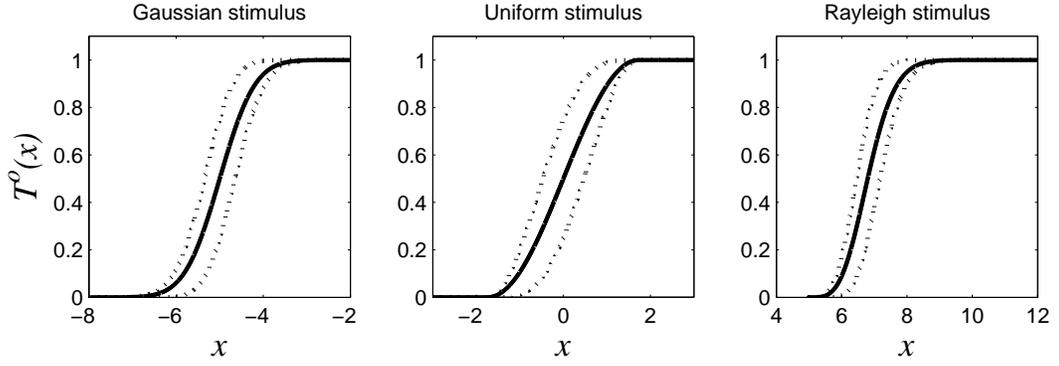}\label{f:meanTuning}}}
\caption{(a) PDFs of three stimulus distributions (solid lines), compared with the derivative of the derived optimal tuning curve ${T^o}'(x)$ (dashed line), and the optimal variability $V^o(x)$ (dotted line). Each has a different mean to illustrate that the mean is not significant. (b) Derived optimal sigmoidal tuning curves (normalized mean firing rate) against stimulus intensity for the three distributions shown in (a). Dotted lines show $T^o(x){\pm} V^o(x)$ (from Eq.~(\ref{P1P})) with $s=1$. This $s$ has been chosen to be very large so that the stimulus dependent variability is clear.}\label{f:Tuning}
\end{center}
\end{figure}

\begin{figure}[t]
\begin{center}
{\subfigure[~Three sigmoidal tuning curves, $T(x)$.]{\includegraphics[scale=0.8]{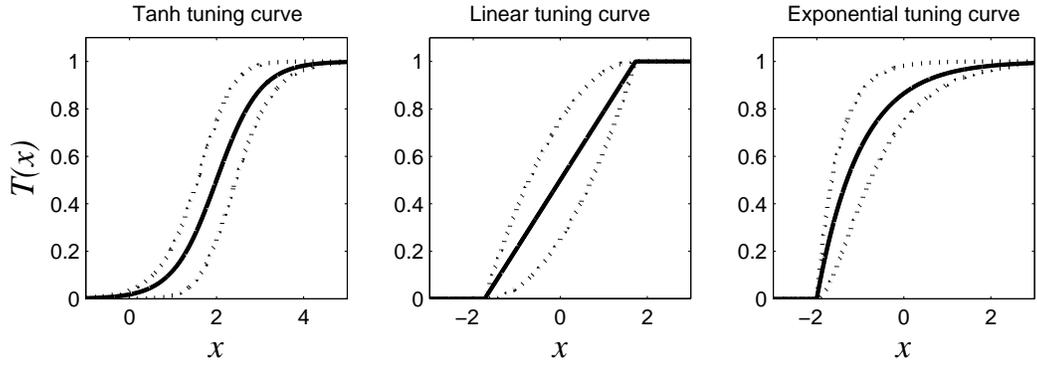}\label{f:meanTuning1}}
\subfigure[~Optimal stimulus PDFs compared to $V(x),~T'(x)$.]{\includegraphics[scale=0.8]{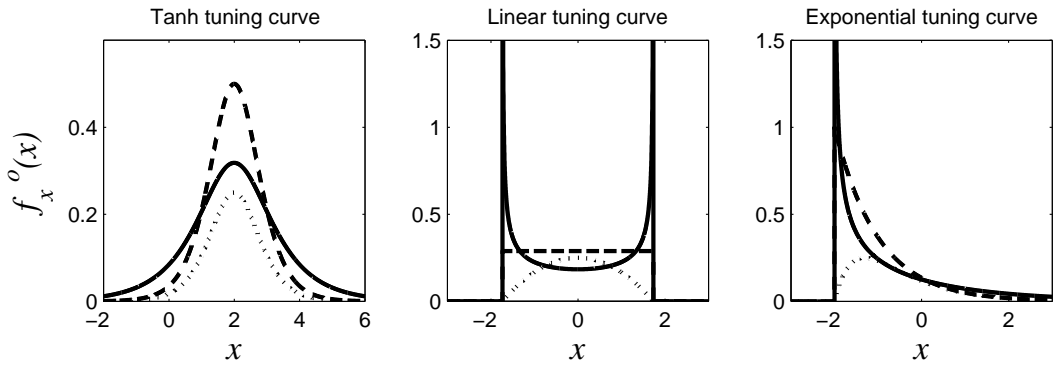}\label{f:varTuning1}}}
\caption{(a) Three sigmoidal tuning curves (normalized mean firing rate) against stimulus intensity. As in Fig.~\ref{f:meanTuning}, dotted lines show $T^o(x){\pm} V^o(x)$ with $s=1$. (b) The optimal stimulus PDF for each tuning curve (solid lines), compared with the derivative of the tuning curve (dashed line), and the variability $V^o(x)$ (dotted line).}\label{f:Tuning1}
\end{center}
\end{figure}

In contrast, optimality conditions for {\em sigmoidal} tuning curves where firing rates increase monotonically with stimulus intensity, as in Eq.~(\ref{hh}) and Figs~\ref{f:meanTuning} and~\ref{f:meanTuning1}, below, have received little attention~\cite{Salinas.06}. The work of~\cite{Bethge.02,Bethge.03a} is a notable exception. The results here differ from~\cite{Bethge.02,Bethge.03a}, in that we maximize {\em mutual information} for sigmoidal tuning curves, rather than optimizing {\em Fisher information}~\cite{Cover2}. Furthermore, our results are far more general than the Poisson assumption of~\cite{Bethge.02,Bethge.03a}, as they apply for Fano-factors other than unity.

\begin{figure}[b]
\begin{center}
{\includegraphics[scale=0.8]{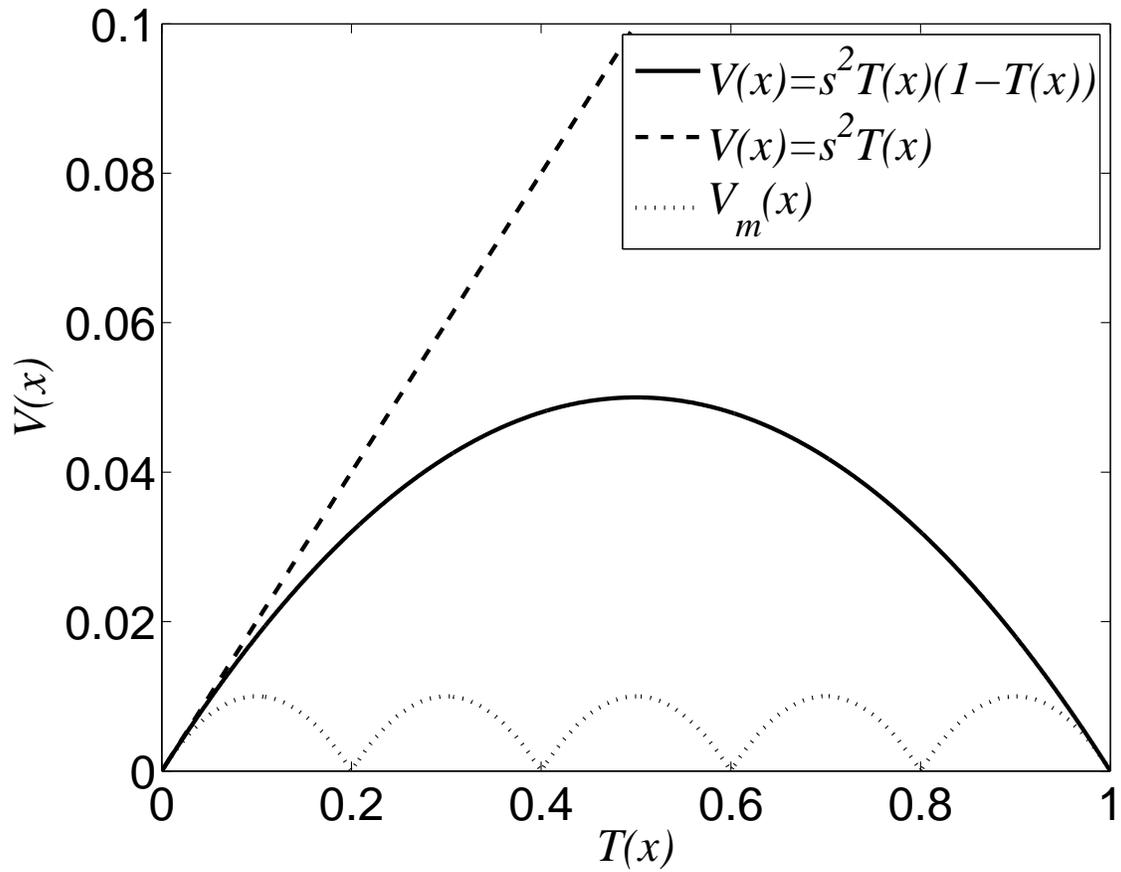}}
\caption{Normalized variance of firing rate $V(x)$ as a function of normalized mean firing rate $T(x)$ for stimulus $x$, and a maximum of $N=5$ spikes: (i) the solid line is the sub-Poisson example considered in this paper, i.e. Eq.~(\ref{P1P}) with $s^2=0.2$; (ii) the dashed line shows the Poisson case where the un-normalized mean is equal to the variance; (iii) the dotted line shows the minimum variance~\cite{deRuyter.97}
~case for $s^2=0.2$.}\label{f:CV}
\end{center}
\end{figure}

Although noisy rate coding neurons are often modeled as a Poisson point process~\cite{Dayan}, in some cases the measured stimulus-dependent variance can be less than the mean (sub-Poisson) or larger than it (super-Poisson)~\cite{Deweese.03}. For example, while the variance typically might be approximately Poisson for firing rates close to zero, it can decrease if the firing rate saturates, due to refractoriness~\cite{Berry.97}
. This can lead, for example, to binomial spiking rather than Poisson spiking~\cite{Bruce.99}
, where the variance is a quadratic function of the mean (as shown in Fig.~\ref{f:CV}, and given below in Eq.~(\ref{P1P})), or even the `scalloped' minimum variance curve~\cite{deRuyter.97}
.

We present our results in terms of {\em normalized} conditional mean firing rate, $T(x)\in[0,1]$ and normalized variance, $V(x)$. Here we consider only monotonically increasing (sigmoidal) tuning curves, so that the derivative of $T(x)$ with respect to stimulus $x$ is strictly nonnegative. Assuming a maximum of $N$ spikes can be produced while a stimulus is unchanged---determined, for example, by refractory times and signal correlation times, or the number of parallel neurons---normalization reduces the mean by a factor of $N$, while the variance is reduced by a factor of $N^2$. Hence a plot of variance against mean for a Poisson system is a straight line with slope $1/N$. Normalized sub-Poisson (i.e. Fano factor smaller than unity) mean-variance curves fall below this line (see Fig.~\ref{f:CV}), and super-Poisson (Fano factor larger than unity) above it.

Our aim is to find optimal tuning curves for the class of sigmoidal neurons or populations where the normalized variance can be expressed as $V(x)=s^2h(T(x))$, where $h(\cdot)$ is an arbitrary function that describes how the variance changes with the mean. The parameter $s^2$ acts to scale the maximum normalized variability, and is typically inversely proportional to $N$, i.e. related to the integration time in an individual neuron, or the number of neurons in a population, as in~\cite{Brunel.98,Bethge.03a}. Our results hold exactly only in the small $s$ limit, meaning that the integration time or number of neurons must be sufficiently large. Otherwise the actual mutual information is closely lower bounded by that of the $s\rightarrow0$ case. Conditional independence in the variability across a population is also assumed, such as that of the cochlear nerve~\cite{Johnson.76}.

Our derivation builds on previous work on the mutual information in neural systems where the {\em instantaneous} normalized firing rate $y$ in response to stimulus value $x$ can be described as~\cite{Brunel.98}
\begin{equation}\label{channel}
    y(x) = T(x)+\sqrt{V(x)}\xi.
\end{equation}
In Eq.~(\ref{channel}),  $T(x)$ and $V(x)=s^2h(T(x))$ are deterministic functions of the stimulus,
and $\xi$ is an arbitrary random variable with zero mean and unit variance.  For the special case where $\xi$ is Gaussian, the result is a {\em
conditionally Gaussian channel}, recently
of much interest in optical and wireless communications~\cite{Chan.05}.

Under regularity conditions on $\xi$,~\cite{Brunel.98} showed that the {\em Fisher information}~\cite{Cover2,Lansky.07} about a specific stimulus value, $x$, in an observation, $y$, for $s$ sufficiently small is
\begin{equation}\label{5:Fisher_I}
    J(x) = \frac{\left(\frac{dT(x)}{dx}\right)^2}{V(x)}k_\xi,
\end{equation}
while the {\em Shannon mutual information}~\cite{Cover2} between the random stimulus and the firing rate is
\begin{align}\label{5:Info_largeN1}
    I(x,y) =H(x)-\frac{1}{2}\int_xf_x(x)\log_2{\left(\frac{2\pi e}{J(x)}\right)}dx+k_\xi^1.
\end{align}
In the above Eqns, $H(x)$ is the differential entropy of the stimulus, $f_x(x)$ is its probability density function (PDF), and $k_\xi$ and $k_\xi^1$ are constants that depend entirely on the PDF of $\xi$. If $\xi$ is Gaussian then $k_\xi=1$ and $k_\xi^1=0$~\cite{Brunel.98}. More general derivations of Eq.~(\ref{5:Info_largeN1}) appear in~\cite{Stein.67,Rissanen.96,Kang.01}.

As discussed in~\cite{Stein.67,Stemmler.96,Brunel.98,McDonnell.07}, the PDF of the stimulus that maximizes the mutual information of Eq.~(\ref{5:Info_largeN1}) is proportional to the square root of the Fisher information. Such a PDF is called {\em Jeffrey's prior}~\cite{Rissanen.96}, which here we denote as $f_J(x)$. Upon letting $k_J=\int_\phi\sqrt{J(\phi)}d\phi$, the optimal stimulus PDF for Eqs~(\ref{channel})--(\ref{5:Info_largeN1}) is therefore
\begin{equation}\label{Jeff}
    f_x^o(x) = f_J(x) :=\frac{\sqrt{J(x)}}{k_J}=\frac{\sqrt{k_\xi}}{sk_J}\frac{\frac{dT(x)}{dx}}{\sqrt{h(T(x))}}.
\end{equation}

What has not previously been recognized is that optimizing Eq.~(\ref{5:Info_largeN1}) can lead to general closed form expressions for the optimal sigmoidal tuning curve, for arbitrary stimulus distributions and non-Poisson variability. This result requires that closed form expressions for the cumulative distribution function (CDF) of the optimal stimulus exist. Using Eqs~(\ref{5:Fisher_I}) and (\ref{Jeff}), this CDF is
\begin{equation}\label{opt_CDF}
    F_X^o(x) = \int_xf_x^o(\phi)d\phi = \frac{\int_0^{T(x)}h(\theta)^{-0.5}d\theta}{\int_0^{1}h(\theta)^{-0.5}d\theta},
\end{equation}
which is independent of $s$ and $\xi$. If Eq.~(\ref{opt_CDF}) can be inverted to isolate $T(x)$ on one side of the equation, the resulting expression also maximizes the mutual information, and is the optimal tuning curve for a given stimulus, $T^o(x)$.

We note that while previous work has discussed the optimal tuning curve for two simple relationships between $T(x)$ and $V(x)$, i.e. constant variance~\cite{Nadal.94}, and the Poisson case~\cite{Brunel.98}, the integrals in Eq.~(\ref{opt_CDF}) are trivial in the former case, and no explicit expression for the optimal tuning curve for arbitrary stimuli was given in the latter.

Although Eq.~(\ref{Jeff}) is known to maximize Eq.~(\ref{5:Info_largeN1}), it has also not been recognized that Eq.~(\ref{5:Info_largeN1}) can be rewritten as
\begin{equation}\label{5:sqrtJ3}
    I(x,y) =0.5\log_2{\left(\frac{k_J^2}{2\pi e}\right)}-D(f_x||f_J)+k_\xi^1,
\end{equation}
where $D(\cdot||\cdot)$ represents the relative entropy (Kullback-Leibler divergence)~\cite{Cover2} between the distributions with PDFs $f_x$ and $f_J$~\cite{footnote}. Since relative entropy is always non-negative, the mutual information is maximized when $f_x=f_J$. As well as a new way of verifying the optimality of Jeffrey's prior, Eq.~(\ref{5:sqrtJ3}) allows calculation of the reduction in mutual information when the tuning curve and the stimulus distribution are not optimally matched.

Another unappreciated consequence of maximizing the mutual information is that regardless of whether the stimulus is optimized for a given sigmoidal tuning curve, or {\it vice versa}, the resulting Fisher information can be written as a
function of the stimulus PDF,
\begin{equation}\label{fisher_P}
    J^o(x) =k_J^2f_x(x)^2.
\end{equation}
It is stated in~\cite{Bethge.02} that
constant Fisher information provides Fisher-optimal neural codes.   From Eq.~(\ref{fisher_P}), the Fisher information at Shannon-optimality is
constant iff the stimulus is uniformly distributed. The discussion in~\cite{Bethge.02} relates to the
mean square error (MSE) between a stimulus and a neural response, rather than the mutual information. We therefore conclude that while a uniform stimulus with the corresponding Shannon-optimal tuning curve will provide the minimum MSE out of all stimulus distributions, that otherwise constant Fisher information and Shannon optimality do not coincide.

Further to this, the Cramer-Rao bound states that the reciprocal of the Fisher information provides a lower bound on achievable conditional
MSE estimates of $x$~\cite{Cover2}. The expected value of this
is a lower bound on the MSE between $x$ and any
estimator for $x$ derived from the mean firing rate $y$. If this lower bound is
asymptotically achievable, e.g. by requiring a large number of observations, or $s\rightarrow 0$, then it is known as the minimum asymptotic square error
(MASE)~\cite{Bethge.02}. From Eq.~(\ref{fisher_P}), the MASE when the stimulus and tuning curve jointly maximize $I(x,y)$ is
\begin{equation}\label{MASE1}
    \mbox{MASE}^o = \int_xf_x(x)\frac{1}{J^o(x)}dx = \frac{1}{k_J^2}\int_x\frac{1}{f_x(x)}dx.
\end{equation}
Clearly, if $f_x(x)$ has long tails, the integral in Eq.~(\ref{MASE1}) may diverge, which indicates the MASE is not achievable by any estimator and that maximizing mutual information and minimizing MASE are not equivalent.

The general observations above are now illustrated and verified for a specific example where the variance and mean are related quadratically as
\begin{equation}\label{P1P}
    V(x)=s^2 T(x)(1-T(x)).
\end{equation}
The integrals in Eq.~(\ref{opt_CDF}) can be solved for this relationship and several examples where it holds have appeared in the experimental neural literature~\cite{Bruce.99}
. We find that $k_J = \pi\sqrt{k_\xi}/s$, and hence the optimal stimulus PDF is
\begin{equation}\label{f_x_opt}
    f_x^o(x) = \frac{T'(x)}{\pi\sqrt{T(x)(1-T(x))}}.
\end{equation}
Integrated and inverting Eq.~(\ref{f_x_opt}) leads to the optimal tuning curve,
\begin{equation}\label{hh}
    T^o(x) = 0.5-0.5\cos{(\pi F_x(x))},
\end{equation}
where $F_x(\cdot)$ is the CDF of the stimulus. The resultant maximum mutual information is
\begin{equation}\label{OptInfo}
    I^o(x,y) =0.5\log_2{\left(\frac{\pi k_\xi}{2es^2}\right)}+k_\xi^1.
\end{equation}
In comparison, for the Poisson case $V(x)=s^2T(x)$, the optimal tuning curve is $T^o(x)=F_x^2(x)$, and the maximum mutual information is reduced by $0.5\log_2{(\pi/2)}$.

Eq.~(\ref{hh}) is plotted for several stimulus distribution examples in Fig.~\ref{f:Tuning}, while Eq.~(\ref{f_x_opt}) is plotted for several tuning curves in Fig.~\ref{f:Tuning1}. The most likely values of the optimal stimulus are not necessary close to the mean. For example, the optimal stimulus for a linear tuning curve has an arcsine distribution, which has a $U$-shaped PDF (Fig.~\ref{f:varTuning1}, middle plot), while for a hyperbolic tangent tuning curve, the optimal PDF is the bell-shaped hyperbolic secent distribution (Fig.~\ref{f:varTuning1}, left-most plot).

\begin{figure}[t]
\begin{center}
{\includegraphics[scale=0.8]{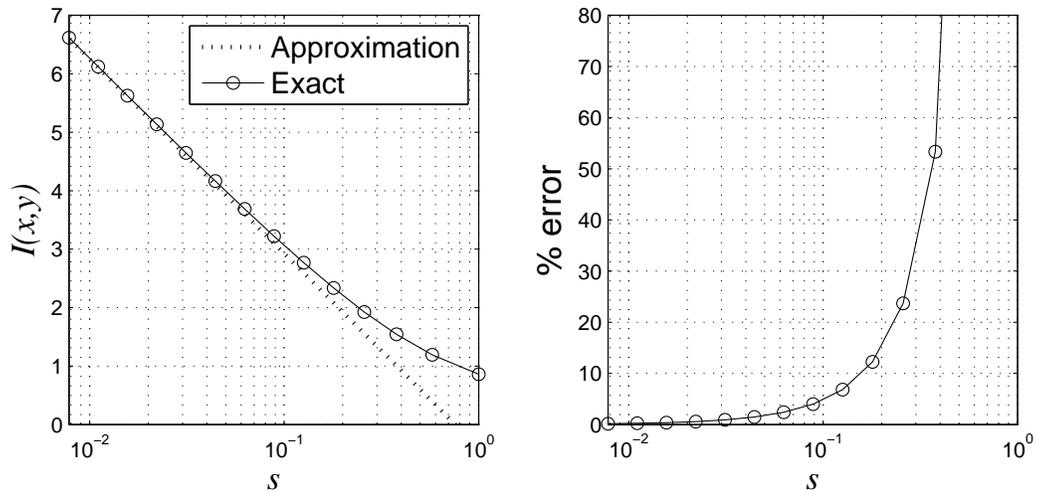}}
\caption{Comparison between the exact mutual information, $I(x,y)$ for Eqs~(\ref{channel}) and~(\ref{P1P}) and the derived mutual information (Eq.~(\ref{OptInfo})) for an optimally matched stimulus and tuning curve, for as a function of $s$.}\label{f:Info}
\end{center}
\end{figure}

From Eq.~(\ref{OptInfo}), the maximum mutual information increases logarithmically with decreasing $s$.  To illustrate the validity of this result for the example of Eq.~(\ref{P1P}), Fig.~\ref{f:Info} shows the exact mutual information calculated numerically for the model of Eq.~(\ref{channel}), as a function of $s$, with the tuning curve and stimulus optimally matched, and $\xi$ Gaussian. Also shown is the mutual information of Eq.~(\ref{OptInfo}), and the percentage error between the two cases. Clearly Eq.~(\ref{OptInfo}) forms a lower bound to the actual mutual information, as discussed in~\cite{Brunel.98}, while the error falls to less than $1\%$ for $s < 0.04$.


We now use Eq.~(\ref{P1P}) to verify our observations about the differences between Shannon and Fisher optimality.  If the stimulus is uniform on $[0,a]$ then the Fisher information is constant. From Eq.~(\ref{fisher_P}), the MASE for the Shannon optimal tuning curve is $\mbox{MASE}_1^o = (as)^2/(\pi^2k_\xi)$. On the other hand, if the tuning curve is $T(x)=x/a$ on $[0,a]$, the Shannon-optimal stimulus has an
arcsine distribution on the same interval, the Fisher information is
non constant, and $\mbox{MASE}_2^o=(as)^2/(8k_\xi)$.
It is clear that $\mbox{MASE}_2^o > \mbox{MASE}_1^o$, which agrees with
constant Fisher information being Shannon optimal only for uniformly distributed stimuli. Indeed, when the stimulus is non-uniformly distributed, different classes of optimal tuning curves to Eq.~(\ref{hh}) might result if the objective was to minimize the MSE instead of maximizing $I(x,y)$.

In closing, if the assumption that $s$ is small is violated, Eq.~(\ref{5:sqrtJ3}) provides a lower bound to the true mutual information achieved for a given stimulus and tuning curve. How different the optimal tuning curve may be for a given stimulus in the event that $s$ is not small is an open question.  Based on preliminary numerical calculations~\cite{Nikitin.unpub}, we conjecture that the optimal tuning curve for $s\sim1$ is composed of a large number of discrete jumps, rather than a smooth increase, which converges to $T^o(x)$ as $s\rightarrow0$. This observation is supported by somewhat related calculations in~\cite{Stein.67,Bethge.03a,Chan.05}.  Future work will address other examples of non-Poisson variability, and consider spontaneous firing and relative refractoriness.

\begin{acknowledgments}
Funding from the Australian Research Council, Post Doctoral Fellowship DP0770747 (McDonnell) and EPSRC grant EP/C523334/1 (Stocks), is gratefully acknowledged. The authors also thank Emilio Salinas and Simon Durrant for valuable discussions.
\end{acknowledgments}


\begin{thebibliography}{24}
\bibitem{footnote}
This relative entropy expression is different from that derived in~\cite{Kang.01}, which is instead that between the conditional response distributions for two different stimulus values.
\end{thebibliography}

\clearpage

\end{document}